\documentstyle[aps,prb,preprint]{revtex}
\begin{document}
\draft \tighten
\title{The rotationally invariant approximation\\
for the two-dimensional $t$-$J$ model}
\author{A.~Sherman}
\address{Institute of Physics, University of Tartu, Riia 142, 51014
Tartu, Estonia}
\author{M.~Schreiber}
\address{Institut f\"ur Physik, Technische Universit\"at, D-09107
Chemnitz, Federal Republic of Germany\\
and School of Engineering and Science, International University Bremen,
Campus Ring 1, D-28759 Bremen, Federal Republic of Germany}
\date{\today}
\maketitle
\begin{abstract}
Using the description in terms of the Hubbard operators hole and spin
Green's functions of the two-dimensional $t$-$J$ model are calculated
in an approximation which retains the rotation symmetry of the spin
susceptibility in the paramagnetic state and has no predefined magnetic
ordering. In this approximation, Green's functions are represented by
continued fractions which are interrupted with the help of the
decoupling corrected by the constraint of zero site magnetization in
the paramagnetic state. Results obtained in this approach for an
undoped 32$\times$32 lattice (the Heisenberg model) and for one hole in
a 4$\times$4 lattice are in good agreement with Monte Carlo and exact
diagonalization data, respectively. In the limit of heavy doping the
hole spectrum described by the obtained formulas acquires features of
the spectrum of weakly correlated excitations.
\end{abstract}

\section{Introduction}
The two-dimensional $t$-$J$ model is one of the most frequently used
models for the description of CuO$_2$ planes of perovskite high-$T_c$
superconductors (for a review, see Ref.~\onlinecite{izyumov}). Together
with the numerical methods -- the exact diagonalization of small
clusters, \cite{dagotto,fehske} Monte Carlo simulations \cite{zhang}
and the density-matrix renormalization-group technique \cite{rommer} --
a number of analytical methods, such as the mean-field slave-boson
\cite{kane} and spin-wave approximations, was used for the
investigation of the model. The latter method which is based on the
spin-wave description of the magnetic excitations was shown to be
remarkably accurate in the case of small hole concentrations and zero
temperature. \cite{marsiglio} This approach was extended to the ranges
of moderate hole concentrations and finite temperatures,
\cite{plakida94} in particular with the use of the spin-wave
approximation modified \cite{takahashi} for short-range order.
\cite{sherman98} The positions, symmetry and size of the pseudogaps in
the hole and magnon spectra, values of the magnetic susceptibility and
spin-lattice relaxation rates obtained in this approach are close to
those observed in photoemission, spin-lattice relaxation and neutron
scattering experiments on cuprate perovskites.
\cite{sherman98,sherman97}

The apparent shortcomings of the spin-wave approximation of the $t$-$J$
model are the violation of the rotation symmetry of the spin
susceptibility components in the paramagnetic state, the predefined
magnetic ordering in the N\'eel state which serves as the reference
state of the approximation, and the neglect of the kinematic
interaction. In this paper we try to overcome these shortcomings by
using the description in terms of Hubbard operators. Green's functions
constructed from these operators are calculated with the use of the
continued fraction representations following from the Mori projection
procedure. \cite{mori} To interrupt these otherwise infinite continued
fractions we use decouplings of the higher-order Green's functions
arising in later stages of this calculation procedure. Following the
idea of Ref.~\onlinecite{kondo}, a correction parameter is introduced
in these decouplings to fulfill the constraint of zero site
magnetization in the paramagnetic state. In this state the obtained
components of the spin Green's functions are rotationally invariant.
The self-energy equations are similar in their form to the equations
derived in the modified spin-wave approximation. \cite{sherman98} In
the case of heavy doping the pole in the hole Green's function
corresponds to a weakly correlated nearest-neighbor band. To check the
validity of the obtained equations in the opposite case of light doping
we have performed calculations for conditions which allow comparison
with exact diagonalization and Monte Carlo results. We found good
agreement of our results with the results of
Refs.~\onlinecite{dagotto,makivic} for spin correlations in an undoped
32$\times$32 lattice and for the hole spectral function of a 4$\times$4
lattice with one hole. To gain a notion of the spectral function in
larger lattices it was calculated in a 20$\times$20 cluster.

\section{Description of the model}
The Hamiltonian of the two-dimensional $t$-$J$ model reads
\begin{equation}
H=\sum_{\bf nm\sigma}t_{\bf nm}a^\dagger_{\bf n\sigma}a_{\bf
m\sigma}+\frac{1}{2}\sum_{\bf nm}J_{\bf nm}\left(s^z_{\bf n}s^z_{\bf
m}+s^{+1}_{\bf n}s^{-1}_{\bf m}\right)+\mu\sum_{\bf n}X_{\bf n},
\label{hamiltonian}\end{equation}
where $a_{\bf n\sigma}=|{\bf n}\sigma\rangle\langle{\bf n}0|$ is the
hole annihilation operator, {\bf n} and {\bf m} label sites of the
square lattice, $\sigma=\pm 1$ is the spin projection, $|{\bf
n}\sigma\rangle$ and $|{\bf n}0\rangle$ are site states corresponding
to the absence and presence of a hole on the site. If Hamiltonian
(\ref{hamiltonian}) is obtained from the extended Hubbard Hamiltonian,
\cite{emery} these states are linear combinations of the products of
the respective $3d_{x^2-y^2}$ copper and $2p_\sigma$ oxygen orbitals.
\cite{sherman98} We take into account nearest neighbor interactions
only, $t_{\bf nm}=t\sum_{\bf a}\delta_{\bf n,m+a}$ and $J_{\bf
nm}=J\sum_{\bf a}\delta_{\bf n,m+a}$ where $t$ and $J$ are hopping and
exchange constants and the four vectors {\bf a} connect nearest
neighbor sites. The spin-$\case{1}{2}$ operators can be written in the
Dirac notations as $s^z_{\bf n}=\case{1}{2}\sum_\sigma\sigma|{\bf
n}\sigma\rangle\langle{\bf n}\sigma|$ and $s^\sigma_{\bf n}=|{\bf
n}\sigma\rangle\langle{\bf n},-\sigma|$. The chemical potential $\mu$
is included into Hamiltonian (\ref{hamiltonian}) to control the hole
concentration. $X_{\bf n}=|{\bf n}0\rangle\langle{\bf n}0|$. The term
$-\case{J}{8}\sum_{\bf na}X_{\bf n}X_{\bf n+a}$ is frequently included
into Hamiltonian (\ref{hamiltonian}). For problems considered below
this term leads to an unessential renormalization of the chemical
potential and therefore it is omitted. The operators $a_{\bf n\sigma}$,
$s^z_{\bf n}$, $s^\sigma_{\bf n}$, and $X_{\bf n}$ are the Hubbard
operators in the space of states of the $t$-$J$ model.

The states $|{\bf n}\sigma\rangle$ and $|{\bf n}0\rangle$ satisfy the
following completeness condition:
\begin{equation}
\sum_\sigma|{\bf n}\sigma\rangle\langle{\bf n}\sigma|+|{\bf
n}0\rangle\langle{\bf n}0|=1.
\label{completeness}\end{equation}
Using this condition and the above expression for $s^z_{\bf n}$ the
constraint of zero site magnetization, which has to be fulfilled in the
paramagnetic state, can be reduced to the form
\begin{equation}
\left\langle s^z_{\bf n}\right\rangle
=\frac{1}{2}\left(1-x\right)-\left\langle s^{-1}_{\bf n}s^{+1}_{\bf
n}\right\rangle=0,
\label{constraint}\end{equation}
where angular brackets denote averaging over the grand canonical
ensemble and the hole concentration $x=\langle X_{\bf n}\rangle$ in the
homogeneous state. It should be noticed that in accord with the
Mermin-Wagner theorem \cite{mermin} the long-range antiferromagnetic
ordering is destroyed for any nonzero temperature in the
two-dimensional system. Therefore the fulfillment of constraint
(\ref{constraint}) has to be ensured for the considered states.

The above operators satisfy the following commutation (anticommutation)
relations:
\begin{eqnarray}
&&\left[s^{-1}_{\bf n},s^{+1}_{\bf m}\right]=-2s^z_{\bf n}\delta_{\bf
 nm},\quad \left[s^\sigma_{\bf n},s^z_{\bf m}\right]=-\sigma
 s^\sigma_{\bf n}\delta_{\bf nm}, \nonumber\\
&&\left[a_{\bf n\sigma},s^z_{\bf m}\right]=-\frac{1}{2}\sigma a_{\bf
 n\sigma}\delta_{\bf nm},\quad \left[a_{\bf n\sigma},s^{\sigma'}_{\bf
 m}\right]=-a_{\bf n,-\sigma}\delta_{\bf nm}\delta_{\sigma,-\sigma'},
 \nonumber\\
&& \label{commutation} \\
&&\left\{a_{\bf n\sigma},a^\dagger_{{\bf
 m},\sigma'}\right\}=\left(1-s^{-\sigma}_{\bf n}s^\sigma_{\bf
 n}\right)\delta_{\bf nm}\delta_{\sigma\sigma'}+s^\sigma_{\bf
 n}\delta_{\bf nm}\delta_{\sigma,-\sigma'},\quad \left\{a_{\bf
 n\sigma},a_{\bf m\sigma'}\right\}=0, \nonumber\\
&&\left[s^{-1}_{\bf n},X_{\bf m}\right]=0,\quad\left[s^z_{\bf n},X_{\bf
 m}\right]=0,\quad \left[a_{\bf n\sigma},X_{\bf m}\right]=a_{\bf
 n\sigma}\delta_{\bf nm}.\nonumber
\end{eqnarray}
Notice that the hole creation and annihilation operators do not satisfy
the fermion anticommutation relations. This is the consequence of the
exclusion of doubly occupied site states due to the strong on-site
repulsion [see Eq.~(\ref{completeness})].

\section{Continued fraction representation of Green's functions}
To investigate the energy spectrum and magnetic properties we shall
calculate the hole and spin retarded Green's functions
\begin{equation}
G({\bf k}t)=\left\langle\!\left\langle a_{\bf k\sigma}\Big|
a^\dagger_{\bf k\sigma}\right\rangle\!\right\rangle_t
=-i\theta(t)\left\langle\!\left\{a_{\bf k\sigma}(t),a^\dagger_{\bf
k\sigma}\right\}\!\right\rangle,\; D({\bf k}t)=\left\langle\!
\left\langle s^z_{\bf k}\Big|s^z_{\bf -k}\right\rangle\!\right
\rangle_t=-i\theta(t)\left\langle\!\left[s^z_{\bf k}(t),s^z_{\bf
-k}\right]\!\right\rangle,
\label{green}\end{equation}
where $a_{\bf k\sigma}=N^{-1/2}\sum_{\bf n}\exp(-i{\bf kn})a_{\bf
n\sigma}$, $s^z_{\bf k}=N^{-1/2}\sum_{\bf n}\exp(-i{\bf kn})s^z_{\bf
n}$, $N$ is the number of sites and $a_{\bf k\sigma}(t)=\exp(iHt)a_{\bf
k\sigma}\exp(-iHt)$. In the considered states $G({\bf k}t)$ does not
depend on $\sigma$.

To calculate the above Green's functions we use their continued
fraction representations which can be obtained using the Mori
projection operator technique. \cite{mori} Let us consider the inner
product $\left|A\cdot B^\dagger\right|$ of the operators $A$ and $B$
which is defined in such a manner that the following conditions are
fulfilled: i) $\left|(aA+bB)\cdot C^\dagger\right|=a\left|A\cdot
C^\dagger\right|+ b\left|B\cdot C^\dagger\right|$, $a$ and $b$ are
arbitrary numbers; ii) $\left|[A,H]\cdot B^\dagger\right|=\left|A
\cdot\left[H,B^\dagger\right]\right|$; iii) $\left|A\cdot
B^\dagger\right|=\left|B\cdot A^\dagger\right|^*$. We notice that the
inner products defined as $\left\langle\left\{A,
B^\dagger\right\}\right\rangle$,
$\left\langle\left[A,B^\dagger\right]\right\rangle$, and
\begin{equation}
\left(A,B^\dagger\right)=i\int_0^\infty dt\, e^{-\eta t}\left\langle
\left[A(t),B^\dagger\right]\right\rangle,\quad \eta\rightarrow +0
\label{innprod}\end{equation}
satisfy the above properties. Let us divide the result of the
commutation of some operator $A_0$ with the Hamiltonian into
longitudinal and transversal parts with respect to $A_0$. The
transversal part $A_1$ is determined as an operator the inner product
of which with $A_0$ is equal to zero. Thus,
\begin{equation}
[A_0,H]=E_0A_0+A_1,
\label{frststp}\end{equation}
where $E_0$ is determined from the condition $\left|A_1\cdot
A_0^\dagger\right|=0$,
$$E_0=\left|[A_0,H]\cdot A_0^\dagger\right|\,\left|A_0\cdot
A_0^\dagger\right|^{-1}.$$
Given $A_0$ and $E_0$, the operator $A_1$ may be found from
Eq.~(\ref{frststp}). The commutator of $A_1$ with the Hamiltonian will
contain already three terms,
$$[A_1,H]=E_1A_1+A_2+F_0A_0.$$
The coefficients $E_1$ and $F_0$ and the new operator $A_2$ are
determined with the use of the two orthogonality conditions
$\left|A_2\cdot A_i^\dagger\right|=0$, $i=0,1$,
$$E_1=\left|[A_1,H]\cdot A_1^\dagger\right|\,\left|A_1\cdot
A_1^\dagger\right|^{-1},\quad F_0=\left|[A_1,H]\cdot
A_0^\dagger\right|\,\left|A_0\cdot A_0^\dagger \right|^{-1}=\left|
A_1\cdot A_1^\dagger\right|\,\left|A_0\cdot A_0^\dagger\right|^{-1},$$
where we have used the properties of the inner product. This procedure
can be continued. In each step of it the coefficients and a new
operator are determined by the conditions of the orthogonality of this
operator to all operators obtained previously. Using the properties of
the inner product it can be shown that only the $n$-th, $(n+1)$-th and
$(n-1)$-th operators appear in the commutator of the $n$-th operator
with the Hamiltonian, \cite{sherman87}
\begin{eqnarray}
&&[A_n,H]=E_nA_n+A_{n+1}+F_{n-1}A_{n-1},\nonumber\\
&&\label{lanczos}\\
&&E_n=\left|[A_n,H]\cdot A_n^\dagger\right|\,\left|A_n\cdot A_n^\dagger
 \right|^{-1},\quad F_{n-1}=\left|A_n\cdot
 A_n^\dagger\right|\,\left|A_{n-1}\cdot A_{n-1}^\dagger\right|^{-1}.
 \nonumber
\end{eqnarray}
As can be seen, algorithm (\ref{lanczos}) is the modification of the
Lanczos orthogonalization procedure which is well known in
computational mathematics (see, e.g., Ref.~\onlinecite{cullum} and
references therein) and physics. \cite{schreiber} In
Eq.~(\ref{lanczos}), operators $A_n$ play the role of mutually
orthogonal wave functions or vectors in the usual Lanczos procedure.

Following Mori \cite{mori} we introduce the projection operator $P_n$
which projects an arbitrary operator $Q$ on the operator $A_n$,
$$P_nQ= \left|Q\cdot A_n^\dagger\right| \,\left|A_n\cdot
A_n^\dagger\right|^{-1}\!\!A_n,$$
and determine the time evolution of the operator $A_n$ by the equation
\begin{equation}
i\frac{d}{dt}{A}_{nt}=\prod_{k=0}^{n-1}(1-P_k)[A_{nt},H],\quad
A_{n,t=0}=A_n,
\label{timevol}\end{equation}
Due to the projection operators in Eq.~(\ref{timevol}) this time
dependence of the operator differs from the conventional one except the
dependence of $A_0$. To underline this difference we use the subscript
notation for the time dependence in Eq.~(\ref{timevol}). Notice also
that in accord with this equation $A_{nt}$ remains orthogonal to
operators $A_i$, $i<n$ for $t>0$. Let us divide $A_{nt}$ into two
parts,
$$A_{nt}=R_n(t)A_n+A'_{nt},\quad  R_n(t)=\left|A_{nt}\cdot
A_n^\dagger\right|\! \left|A_n\cdot A_n^\dagger\right|^{-1}.$$
From this definition it follows that $A'_{nt}=(1-P_n)A_{nt}$. Equations
(\ref{lanczos}) and (\ref{timevol}) determine the time evolution of
this operator,
$$i\frac{d}{dt}{A}'_{nt}=R_n(t)A_{n+1}+\prod_{k=0}^n(1-P_k)[A'_{nt},H].$$
Solving this equation we find
$$A_{nt}=R_n(t)A_n-i\int_0^td\tau R_n(\tau)A_{n+1,t-\tau}.$$
This result allows us to obtain the following equation for the
functions $R_n(t)$:
$$i\frac{d}{dt}R_n(t)=E_nR_n(t)-iF_n\int_0^td\tau
R_n(\tau)R_{n+1}(t-\tau).$$
After the Laplace transformation $R_n(\omega)=-i\int_0^\infty
dt\,\exp(i\omega t)R_n(t)$ this equation reads
\begin{equation}
R_n(\omega)=\left[\omega-E_n-F_nR_{n+1}(\omega)\right]^{-1}.
\label{recursive}\end{equation}

If the inner product is defined as the average of the commutator
(anticommutator) of operators, the function
$\widetilde{R}_0(\omega)=R_0(\omega)\left|A_0\cdot A_0^\dagger\right|$
coincides with the Fourier transform, $\left\langle\!\left\langle
A_0\Big|A_0^\dagger
\right\rangle\!\right\rangle_\omega=\int_{-\infty}^\infty
dt\exp(i\omega t)\left\langle\! \left\langle
A_0\Big|A_0^\dagger\right\rangle\!\right\rangle_t$, of the commutator
(anticommutator) retarded Green's functions of the type of
Eq.~(\ref{green}). If the inner product is defined by
Eq.~(\ref{innprod}), the function $\widetilde{R}_0(\omega)$ coincides
with Kubo's relaxation function
\begin{equation}
\left(\!\left(A_0\Big|A_0^\dagger\right)\!\right)_\omega=
\int^\infty_{-\infty}dt\,e^{i \omega
t}\left(\!\left(A_0\Big|A_0^\dagger\right)\!\right)_t,\quad
\left(\!\left(A_0\Big|A_0^\dagger\right)\!\right)_t=\theta(t)\int^\infty_t
dt'\left\langle \left[A_0(t'),A_0^\dagger\right]\right\rangle.
\label{kubo}\end{equation}

From Eq.~(\ref{recursive}) for all these functions we obtain the
following continued fraction representation:
\begin{equation}
\widetilde{R}_0(\omega)=\frac{\displaystyle\left|A_0\cdot A_0^\dagger
\right|}{\displaystyle \omega-E_0-\frac{\displaystyle
F_0}{\displaystyle\omega-E_1-\frac{\displaystyle F_1}{\ddots}}}.
\label{cfraction}\end{equation}
Thus the recursive procedure (\ref{lanczos}) in course of which the
coefficients $E_n$ and $F_n$ of the continued fraction
(\ref{cfraction}) are determined allows us to calculate Green's or
Kubo's relaxation functions.

\section{The spin Green's function}
The direct application of Eqs.~(\ref{lanczos}) and (\ref{cfraction}) to
the spin Green's function $D({\bf k}\omega)$, Eq.~(\ref{green}), meets
with difficulties because the inner product $\left\langle\left[s^z_{\bf
k},s^z_{\bf -k}\right]\right\rangle$ in the numerator of the continued
fraction (\ref{cfraction}) is equal to zero. To overcome this
difficulty we consider Kubo's relaxation function
$\left(\!\left(s^z_{\bf k}\Big|s^z_{\bf -k}\right)\! \right)$ defined
in Eq.~(\ref{kubo}). In this case the inner product (\ref{innprod}) in
the numerator of the respective continued fraction is nonzero. After
calculating the relaxation function the spin Green's function can be
obtained from the relation
\begin{equation}
\omega\left(\!\left(s^z_{\bf k}\Big|s^z_{\bf
-k}\right)\!\right)=\left\langle\!\left\langle s^z_{\bf k}\Big|s^z_{\bf
-k}\right\rangle\!\right\rangle+\left(s^z_{\bf k},s^z_{\bf -k}\right),
\label{kg}\end{equation}
where we dropped the subscript $\omega$ in the relaxation and Green's
functions.

We postpone the calculation of the numerator $\left(s^z_{\bf
k},s^z_{\bf -k}\right)$ of the continued fraction and consider its
other coefficients. From definition (\ref{innprod}) we find that
$E_0\left( s^z_{\bf k},s^z_{\bf -k}\right)=\left(i\dot{s}^z_{\bf
k},s^z_{\bf -k} \right)=\left\langle\left[s^z_{\bf k},s^z_{\bf
-k}\right]\right\rangle= 0$ and therefore $A_1$ is the Fourier
transform of the operator
\begin{equation}
i\dot{s}^z_{\bf l}=\frac{1}{2}\sum_{\bf mn}J_{\bf mn}\left(\delta_{\bf
ln}-\delta_{\bf lm}\right)s^{+1}_{\bf n}s^{-1}_{\bf
m}+\frac{1}{2}\sum_{\bf mn\sigma}t_{\bf mn}\sigma\left( \delta_{\bf
lm}-\delta_{\bf ln}\right)a^\dagger_{\bf n\sigma}a_{\bf m\sigma}={\cal
A}^s_{\bf l}+{\cal A}^h_{\bf l}.
\label{sder}\end{equation}
Here the dot over the operator indicates the time derivative. As can be
seen, $A_1$ contains contributions from spin and hole components ${\cal
A}^s$ and ${\cal A}^h$. Using this result in calculating $R_1(\omega)$
which is the Laplace transform of the function
$\left(A_{1t},A_1^\dagger\right)$ we neglect the terms $\left({\cal
A}_{1t}^h,{\cal A}^{s\dagger}\right)$ and $\left({\cal A}_{1t}^s,{\cal
A}^{h\dagger}\right)$. This approximation is motivated by vanishing
values of these correlations obtained with the decoupling. Therefore
\begin{equation}
\left(A_{1t},A_1^\dagger\right)\approx\left({\cal A}^h(t),{\cal A}^{h
\dagger}\right)+\left({\cal A}^s_{1t},{\cal A}^{s\dagger}\right),
\label{appr}\end{equation}
where we have additionally neglected the difference between ${\cal
A}^h_{1t}$ and ${\cal A}^h(t)$ (again due to zero values of the
respective decoupling). In accord with our estimation the influence of
terms connected with holes in $\left(s^z_{\bf k},s^z_{\bf -k}\right)$
and $\left(\!\left({\cal A}^s_{\bf k}\Big|{\cal A}^{s\dagger}_{\bf
k}\right)\!\right)$ on the spin Green's function is small in comparison
with the quantity $\left(\!\left({\cal A}^h_{\bf k}\Big|{\cal
A}^{h\dagger}_{\bf k}\right)\!\right)$ even for moderate hole
concentrations. Therefore in the forthcoming discussion we neglect
these terms in $\left(s^z_{\bf k},s^z_{\bf -k}\right)$ and
$\left(\!\left({\cal A}^s_{\bf k}\Big|{\cal A}^{s\dagger}_{\bf
k}\right)\!\right)$ and consider the time evolution of operators in
these quantities as determined solely by the Heisenberg part of
Hamiltonian (\ref{hamiltonian}). In this approximation the numerator of
the continued fraction representing $\left(\!\left({\cal A}^s_{\bf
k}\Big|{\cal A}^{s\dagger}_{\bf k}\right)\!\right)$ reads
\begin{equation}
\left({\cal A}^s_{\bf k},{\cal A}^{s\dagger}_{\bf
k}\right)=\left(i\dot{s}^z_{\bf k},-i\dot{s}^z_{\bf -k}\right)=\left
\langle\left[i\dot{s}^z_{\bf k},s^z_{\bf -k}\right]\right\rangle=
4JC_1(\gamma_{\bf k}-1),
\label{nf}\end{equation}
where $\gamma_{\bf k}=\case{1}{4}\sum_{\bf a}\exp(i{\bf ka})$, $C_p=
\case{1}{N}\sum_{\bf k}\gamma^p_{\bf k}C_{\bf k}$ and $C_{\bf k}=
\sum_{\bf n}\exp[i{\bf k(n-m)}]\langle s^{+1}_{\bf n}s^{-1}_{\bf m}
\rangle$. For $E_1$ we get $E_1\left(i\dot{s}^z_{\bf
k},-i\dot{s}^z_{\bf -k}\right)=\left(i^2\ddot{s}^z_{\bf
k},-i\dot{s}^z_{\bf -k}\right)=\left \langle\left[i\dot{s}^z_{\bf
k},-i\dot{s}^z_{\bf -k}\right]\right\rangle =0$. Thus, breaking off the
continued fraction on this step we obtain from Eqs.~(\ref{cfraction}),
(\ref{kg}), (\ref{appr}) and (\ref{nf})
\begin{equation}
D({\bf k}\omega)=\frac{\omega\left(\!\left({\cal A}^h_{\bf k}\Big|{\cal
A}^{h\dagger}_{\bf k}\right)\!\right)+4JC_1(\gamma_{\bf
k}-1)}{\omega^2-2\omega\,\Pi({\bf k}\omega)- \omega^2_{\bf k}},
\label{sgf}\end{equation}
where the polarization operator and the excitation frequency are given
by
\begin{equation}
\Pi({\bf k}\omega)=\frac{1}{2}\left(\!\left({\cal A}^h_{\bf
k}\Big|{\cal A}^{h\dagger}_{\bf k}\right)\!\right)\left(s^z_{\bf
k},s^z_{\bf -k}\right)^{-1}, \quad \omega^2_{\bf k}=4JC_1(\gamma_{\bf
k}-1)\left(s^z_{\bf k},s^z_{\bf -k}\right)^{-1}.
\label{pf}\end{equation}

To calculate $\left(s^z_{\bf k},s^z_{\bf -k}\right)$ in the above
formulas we notice that in the considered case $A_2=i^2\ddot{s}^z_{\bf
k}-\left(i\dot{s}^z_{\bf k},-i\dot{s}^z_{\bf -k}\right)\left(s^z_{\bf
k},s^z_{\bf -k}\right)^{-1}\!s^z_{\bf k}$ and
\begin{equation}
\left(A_2,A_2^\dagger\right)=\left\langle\left[i^2\ddot{s}^z_{\bf
k},-i\dot{s}^z_{\bf
-k}\right]\right\rangle-\frac{16J^2C_1^2(\gamma_{\bf
k}-1)^2}{\left(s^z_{\bf k},s^z_{\bf -k}\right)}=0.
\label{tsr}\end{equation}
We set the above result equal to zero in conformity with the
approximation made above in the continued fraction where we dropped all
terms containing $A_2$ and operators of higher orders. Equation
(\ref{tsr}) can be used for calculating $\left(s^z_{\bf k},s^z_{\bf
-k}\right)$ if the value of $\left\langle\left[i^2\ddot{s}^z_{\bf k},-i
\dot{s}^z_{\bf -k}\right]\right\rangle$ is known. An analogous equation
was obtained in Ref.~\onlinecite{tserkovnikov82} with another method.

We calculate $\left\langle\left[i^2\ddot{s}^z_{\bf k},-i\dot{s}^z_{\bf
-k}\right]\right \rangle$ in Eq.~(\ref{tsr}) by decoupling terms in the
second derivative of $s^z$,
\begin{eqnarray}
i\ddot{s}^z_{\bf l}=\frac{1}{2}\sum_{\bf mn}\Big[
&&J_{\bf lm}J_{\bf ln}\Big(2s^z_{\bf l}s^{+1}_{\bf n}s^{-1}_{\bf
 m}-s^z_{\bf n}s^{+1}_{\bf l}s^{-1}_{\bf m}-s^{+1}_{\bf n}s^z_{\bf
 m}s^{-1}_{\bf l}\Big)+ \nonumber\\
&&J_{\bf lm}J_{\bf mn}\Big(s^z_{\bf n}s^{+1}_{\bf m}s^{-1}_{\bf
 l}-s^z_{\bf m}s^{+1}_{\bf n}s^{-1}_{\bf l}+s^{+1}_{\bf l}s^z_{\bf
 n}s^{-1}_{\bf m}-s^{+1}_{\bf l}s^z_{\bf m}s^{-1}_{\bf n}\Big)\Big].
\label{ssd}\end{eqnarray}
In the decoupling we approximate $s^z_{\bf l}s^{+1}_{\bf n}s^{-1}_{\bf
m}$ by the value $\left[\alpha C_{\bf nm}\left(1-\delta_{\bf
nm}\right)+\case{1}{2}\delta_{\bf nm} \right]s^z_{\bf l}$ where $C_{\bf
nm}= \langle s^{+1}_{\bf n}s^{-1}_{\bf m}\rangle$. In the last
expression we took into account that in accordance with
Eq.~(\ref{constraint}) $C_{\bf nn}=\case{1}{2}$ for $x=0$ [let us
remind that we neglect the influence of holes on the value of
$\left(s^z_{\bf k},s^z_{\bf -k}\right)$]. Following
Ref.~\onlinecite{kondo} the parameter $\alpha$ is introduced to fulfill
the constraint of zero site magnetization (\ref{constraint}) in the
paramagnetic state. Before carrying out the decoupling it has to be
taken into account that terms of Eq.~(\ref{ssd}) in which the site
index of the $s^z$ operator coincides with the site index of $s^{+1}$
or $s^{-1}$ operators cancel each other. To verify this statement it is
necessary to take into consideration that in these terms the operators
$s^z_{\bf n}$ can be substituted by $-\case{1}{2}$, since for the
spin-$\case{1}{2}$ case $s^z_{\bf n}=-\case{1}{2}+s^{+1}_{\bf
n}s^{-1}_{\bf n}$ and $s^{+1}_{\bf n}s^{+1}_{\bf n}=0$. To retain this
exact cancellation it has to be taken into account before the
decoupling. As the result we find
\begin{eqnarray*}
i^2\ddot{s}^z_{\bf l}=
&&\alpha\sum_{\bf mn}\Big[J_{\bf lm}J_{\bf ln}\Big(C_{\bf mn}s^z_{\bf
 l}-C_{\bf lm}s^z_{\bf n}\Big)+J_{\bf ln}J_{\bf mn}\Big(C_{\bf
 ln}s^z_{\bf m}-C_{\bf lm} s^z_{\bf n}\Big)\Big]\\
&&+\sum_{\bf n}J^2_{\bf ln}\Big[(1-\alpha)C_{\bf nn}\Big(s^z_{\bf
 l}-s^z_{\bf n}\Big)+\alpha C_{\bf ln}\Big(s^z_{\bf n}-s^z_{\bf
 l}\Big)\Big],
\end{eqnarray*}
and after the Fourier transformation
\[ i^2\ddot{s}^z_{\bf k}=\omega^2_{\bf k}s^z_{\bf k}, \]
where
\begin{eqnarray}
&&\omega^2_{\bf k}=\frac{\left\langle\left[i^2\ddot{s}^z_{\bf
 k},-i\dot{s}^z_{\bf -k}\right] \right\rangle}{4JC_1(\gamma_{\bf
 k}-1)}=16J^2\alpha|C_1|\left(1-\gamma_{\bf k}\right)\left(
 \Delta+1+\gamma_{\bf k}\right), \nonumber\\
&&\label{freq} \\
&&\Delta=\frac{C_2}{|C_1|}+\frac{1-\alpha}{8\alpha|C_1|}-\frac{3}{4}.
 \nonumber
\end{eqnarray}
Combining Eqs.~(\ref{nf}), (\ref{pf}) and (\ref{freq}) we find
\begin{equation}
(s^z_{\bf k},s^z_{\bf -k})^{-1}=4J\alpha\left(\Delta+1+\gamma_{\bf
k}\right).
\label{inum}\end{equation}

In the absence of holes Eqs.~(\ref{sgf}) and (\ref{freq}) are close to
the equations for the spin Green's function and the excitation
frequency obtained for the two-dimensional Heisenberg antiferromagnet
in Ref.~\onlinecite{shimahara,winterfeldt} with the use of the
equations of motion for Green's functions and Tserkovnikov's formalism,
\cite{tserkovnikov82} respectively. In these works, somewhat more
complicated decouplings were used. These decouplings contain several
decoupling parameters of the type of $\alpha$ which depend on the site
indices in the decoupled average. These additional parameters allow one
to obtain somewhat better agreement with numeric simulations. However,
to fix the additional parameters exterior data from numerical
simulations or the spin-wave theory have to be engaged and the theory
ceases to be closed.

As can be shown by the analogous calculation of the transversal spin
Green's function $\left\langle\!\left\langle s_{\bf k}^{-1}\Big|s_{\bf
k}^{+1}\right\rangle\!\right\rangle$, in the paramagnetic state
\begin{equation}
\left\langle\!\left\langle s_{\bf k}^{-1}\Big|s_{\bf
k}^{+1}\right\rangle\!\right\rangle =2\left\langle\!\left\langle s_{\bf
k}^z\Big|s_{\bf -k}^z\right\rangle\!\right\rangle.
\label{ssc}\end{equation}
Thus, the rotation symmetry of the components of the magnetic
susceptibility is retained in this approach. This fact can be used for
the calculation of parameters $C_1$, $C_2$ and $\alpha$ in the above
formulas. From Eq.~(\ref{sgf}) simplified for the absence of holes,
Eq.~(\ref{ssc}) and the relation
\begin{equation}
\langle s^z_{\bf k}(t)s^z_{-\bf
k}\rangle=\int^\infty_{-\infty}d\omega{\rm e}^{-i\omega t}{\rm
e}^{\beta\omega}n_B(\omega)B({\bf k}\omega),
\label{ssr}\end{equation}
we find
\begin{equation}
C_{\bf k}=4J|C_1|\left(1-\gamma_{\bf k}\right)\omega_{\bf
k}^{-1}\coth\left(\frac{1}{2} \beta\omega_{\bf k}\right),
\label{ck}\end{equation}
where $B({\bf k}E)=-\pi^{-1}{\rm Im}\, D({\bf k}E)$ is the spin
spectral function, $n_B(E)=[\exp(\beta E)-1]^{-1}$ and $\beta=T^{-1}$
is the inverse temperature. Substituting this equation in the
definitions of $C_1$, $C_2$ and in constraint (\ref{constraint}) we
obtain three equations for the three unknown parameters $C_1$, $C_2$
and $\alpha$. This problem can be reduced to the optimization problem
and solved by the steepest descent method.

To check the validity of the approximations made above we used the
obtained formulas for calculating spin correlations in an undoped
antiferromagnet. In Fig.~\ref{figi} our results obtained in a
32$\times$32 lattice for three temperatures are compared with data
of Monte Carlo simulations performed for the same lattice in
Ref.~\onlinecite{makivic}. As can be seen, the agreement is good.
However, it should be noted that at elevated temperatures in our
approximation the spin correlations are systematically
overestimated in comparison with the Monte Carlo results.

As follows from Eq.~(\ref{freq}), for low temperatures and large
crystals the spectrum of elementary spin excitations is close to the
spectrum of spin waves. \cite{manousakis} For an infinite crystal and
$T=0$ we found $\alpha=1.70494$ and $C_2=-C_1= 0.206734$. In this case
in Eq.~(\ref{freq}) the parameter $\Delta=0$ and the excitation
frequency vanishes in the two points of the Brillouin zone, ${\bf
k}=(0,0)$ and $(\pi,\pi)$ (here and below the intersite distance is
taken as the unit of length). For any nonzero temperature $\Delta$
becomes finite which generates a gap at the $(\pi,\pi)$ point. It can
be shown \cite{takahashi,shimahara} that the gap leads to the
exponential decay of spin correlations with distance and the respective
correlation length is defined by the magnitude of the gap. Thus, in
agreement with the Mermin-Wagner theorem \cite{mermin} for a nonzero
temperature the long-range antiferromagnetic order is destroyed in the
considered two-dimensional system.

Now let us calculate the polarization operator $\Pi({\bf k}\omega)$,
Eq.~(\ref{pf}). Using the decoupling which is equivalent to the Born
approximation \cite{plakida94} and the relations
\begin{equation}
\langle a_{\bf k\sigma}(t)a^\dagger_{\bf
k\sigma}\rangle=\int^\infty_{-\infty}d\omega{\rm e}^{-i\omega t}{\rm
e}^{\beta\omega}n_F(\omega)A({\bf k}\omega),\quad \langle
a^\dagger_{\bf k\sigma}a_{\bf
k\sigma}(t)\rangle=\int^\infty_{-\infty}d\omega{\rm e}^{-i\omega
t}n_F(\omega)A({\bf k}\omega),
\label{hsr}\end{equation}
we find
\begin{eqnarray}
&&{\rm Im}\,\Pi({\bf k}\omega)=\frac{\pi}{\omega}\sum_{\bf k'}f^2_{\bf
 k'k}\int^\infty_{-\infty}d\omega'\left[n_F(\omega')-n_F(\omega'-\omega)
 \right]A({\bf k'-k},\omega'-\omega)A({\bf k'}\omega'), \nonumber\\
&&\label{po}\\
&&{\rm Re}\,\Pi({\bf k}\omega)={\cal
 P}\int^\infty_{-\infty}\frac{d\omega'}{\pi} \frac{{\rm Im}\,\Pi({\bf
 k}\omega')}{\omega'-\omega},\nonumber
\end{eqnarray}
where $A({\bf k}\omega)=-\pi^{-1}{\rm Im}\,G({\bf k}\omega)$ is the
hole spectral function, $n_F(\omega)=[\exp(\beta\omega)+1]^{-1}$,
$f_{\bf k'k}= 2tN^{-1/2}( \gamma_{\bf k'}-\gamma_{\bf k'-k})(s^z_{\bf
k},s^z_{\bf -k})^{-1/2}$ and ${\cal P}$ indicates Cauchy's principal
value of the integral. We notice that Eq.~(\ref{po}) is close in its
form to the polarization operator obtained for the $t$-$J$ model in the
spin-wave approximation. \cite{plakida94,sherman98} We cannot directly
compare the interaction constants, because the definition of the spin
Green's function in this paper differs from the magnon Green's
functions in Refs.~\onlinecite{plakida94,sherman98}. However, we notice
that the spin-wave interaction constant and the respective quantity
$f^2_{\bf k'k}\omega^{-1}_{\bf k}$ in Eq.~(\ref{po}) are of the same
order of magnitude and tend to zero linearly with $|{\bf k}|$ when
$|{\bf k}|\rightarrow 0$. The spin-wave constant behaves analogously
near the $(\pi,\pi)$ point, while the quantity $f^2_{\bf
k'k}\omega^{-1}_{\bf k}$ does so only in the case of an infinite
crystal and zero temperature.

\section{The hole Green's function}
Now let us consider the hole Green's function. To use the continued
fraction representation (\ref{cfraction}) for the anticommutator
Green's function $G({\bf k}t)$, Eq.~(\ref{green}), the average of the
anticommutator of operators has to be taken as the definition of the
inner product in the recursive procedure (\ref{lanczos}). From the
commutation relations (\ref{commutation}) we find for the numerator of
the continued fraction $\left\langle\left\{a_{\bf
k\sigma},a^\dagger_{\bf k\sigma}\right\}
\right\rangle=\case{1}{2}(1+x)=\phi$ and for the time derivative
\begin{equation}
i\dot{a}_{\bf l\sigma}=\sum_{\bf m}t_{\bf lm}\left[\left(1-
 s^{-\sigma}_{\bf l}s^\sigma_{\bf l}\right)s^\sigma_{\bf m}+
 s^\sigma_{\bf l}\right]a_{\bf m,-\sigma}-\frac{1}{2}\sum_{\bf m}
 J_{\bf lm}\left(\sigma s^z_{\bf m}s^\sigma_{\bf l}+s^\sigma_{\bf m}
 \right)a_{\bf l,-\sigma}+\mu a_{\bf l\sigma}.
\label{ader}\end{equation}
With these results we get
\begin{eqnarray}
&&E_0=\left\langle\!\left\{i\dot{a}_{\bf k\sigma},a^\dagger_{\bf
 k\sigma}\right\}\!\right\rangle\left\langle\!\left\{a_{\bf k\sigma},
 a^\dagger_{\bf k\sigma}\right\}\!\right\rangle^{-1}=\varepsilon_{\bf
 k}+\mu', \nonumber\\
&&\label{urf}\\
&&\varepsilon_{\bf k}=(4t\phi+6tC_1\phi^{-1}-3JF_1\phi^{-1})
 \gamma_{\bf k}, \quad
 \mu'=\mu+(4tF_1-3JC_1)\phi^{-1}, \nonumber
\end{eqnarray}
where $F_1=N^{-1}\sum_{\bf k}\gamma_{\bf k}F_{\bf k}$ and $F_{\bf k}=
\sum_{\bf n}\exp[i{\bf k(n-m)}]\left\langle a^\dagger_{\bf n}a_{\bf m}
\right\rangle$.

The estimation of $t$ and $J$ based on the parameters of the extended
Hubbard model \cite{mcmahan} gives $J/t$ lying in the range $0.2-0.3$.
For low hole concentrations we can approximate the parameter $C_1$ by
its value in an undoped lattice. For $T=0.02t$ in a 4$\times$4 lattice
$C_1=0.2119$, while in a 20$\times$20 lattice $C_1=0.2068$. With these
parameters the unrenormalized hole dispersion can be estimated as
$\varepsilon_{\bf k}\approx -0.27t\gamma_{\bf k}$ in the former case
and $-0.47t\gamma_{\bf k}$ in the latter case. Thus the first
approximation of the recursive procedure describes a band which is much
narrower than the two-dimensional nearest-neighbor band in the absence
of correlations $4t\gamma_{\bf k}$. The reason for this is the
antiferromagnetic alignment of spins when the hole movement is
accompanied by the spin flipping. With increasing the hole
concentration $C_1\rightarrow 0$ and the unrenormalized dispersion
tends to its uncorrelated value.

The hole Green's function reads
\begin{eqnarray}
&&G({\bf k}\omega)=\frac{\phi}{\omega-\varepsilon_{\bf k}-\mu'-
 \Sigma({\bf k}\omega)},\nonumber\\
&&\label{hgf}\\
&&\Sigma({\bf k}\omega)=\phi^{-1}\left\langle\!\left\langle A_1\Big|
 A_1^\dagger\right\rangle\!\right\rangle,\quad A_1=i\dot{a}_{\bf k
 \sigma}-\left(\varepsilon_{\bf k}+\mu'\right)a_{\bf k\sigma},\nonumber
\end{eqnarray}
where the difference between $A_{1t}$ and $A_1(t)$ was neglected. Due
to the mentioned smallness of $\varepsilon_{\bf k}$ for low hole
concentrations and of $J$ in comparison with $t$, only the term
$N^{-1/2}\sum_{\bf lm}\exp(-i{\bf kl})\,t_{\bf
lm}\left[\left(1-s^{-\sigma}_{\bf l}s^\sigma_{\bf
l}\right)s^\sigma_{\bf m}+s^\sigma_{\bf l}\right]a_{\bf m,-\sigma}$ may
be retained in $A_1$ in the calculation of $\left\langle\!\left\langle
A_1\Big|A_1^\dagger \right\rangle\!\right\rangle$. The terms in $A_1$
which are linear in spin operators produce the following contribution
to the self-energy:
$$\frac{32t^2}{N\phi}\sum_{\bf k'}\int\!\!\!\int^\infty_{-\infty}
d\omega_1d\omega_2\frac{n_F(-\omega_1)+n_B(\omega_2)}{\omega-\omega_1-
\omega_2+i\eta}(\gamma_{\bf k}+ \gamma_{\bf k-k'})^2A({\bf
k-k'},\omega_1)B({\bf k'},\omega_2).$$
Up to the prefactor this
expression coincides with the respective term in the hole
self-energy calculated in the spin-wave approximation.
\cite{sherman98}

The term with three spin operators in $A_1$ produces terms in the
self-energy which contain two- and three-spin Green's functions. To
calculate these functions one would have to solve the respective
self-energy equations which could be derived in the same way as the
equations in the previous section. However, such program would
essentially complicate the calculation procedure. One of the possible
ways to overcome this difficulty is to use the decoupling in the same
manner as we applied it in the previous section, this time in the term
with three spin operators in $A_1$. However, the comparison with the
exact diagonalization data shows that this approximation does not give
satisfactory results. Another way of simplification is suggested by the
above observation that the terms with the one-spin Green's function in
self-energy (\ref{hgf}) are similar to the terms obtained in the
spin-wave approximation. This gives grounds to suppose that the
correcting terms containing two- and three-spin Green's functions can
be approximated by the respective terms of the spin-wave approximation
modified for short-range antiferromagnetic order. Using the results of
Refs.~\onlinecite{sherman98,sherman99} and Eqs.~(\ref{ssr}),
(\ref{hsr}) we find
\begin{eqnarray}
{\rm Im}\,\Sigma({\bf k}\omega)
&=&\frac{16\pi t^2}{N\phi}\sum_{\bf k'} \int^\infty_{-\infty}
 d\omega'\left[n_B(-\omega')+n_F(\omega-\omega')\right]\sqrt{\frac{1+
 \gamma_{\bf k'}}{\Delta +1-\gamma_{\bf k'}}}\nonumber\\
&\times&\left[\left(\gamma_{\bf k-k'}+\gamma_{\bf
 k}\right)\sqrt[4]{\frac{1-\gamma_{\bf k'}}{\Delta +1+\gamma_{\bf
 k'}}}+{\rm sgn}(\omega')\left(\gamma_{\bf k-k'}-\gamma_{\bf
 k}\right)\sqrt[4]{\frac{1+\gamma_{\bf k'}}{\Delta +1-\gamma_{\bf
 k'}}}\right]^2\nonumber\\
&&\label{se}\\
&\times&A({\bf k-k'},\omega-\omega')B({\bf k'}\omega'),\nonumber\\
{\rm Re}\,\Sigma({\bf k}\omega)
&=&{\cal P}\int^\infty_{-\infty}\frac{d\omega'}{\pi} \frac{{\rm
 Im}\,\Sigma({\bf k}\omega')}{\omega'-\omega}.\nonumber
\end{eqnarray}

For low hole concentrations the spin spectral function in the above
equation can be substituted by its value in the absence of holes,
\begin{equation}
B({\bf k}\omega)=\frac{1}{2}\sqrt{\frac{|C_1|}{\alpha}}\sqrt{
\frac{1-\gamma_{\bf k}}{\Delta+1+\gamma_{\bf
k}}}\left[\delta\left(\omega-\omega_{\bf k}\right)-
\delta\left(\omega+\omega_{\bf k}\right)\right].
\label{ssf}\end{equation}
With this substitution and for low temperatures Eq.~(\ref{se}) acquires
the form
\begin{eqnarray}
{\rm Im}\,\Sigma({\bf k}\omega)
&=&-\frac{8\pi t^2}{N\phi}\sqrt{\frac{|C_1|}{\alpha}} \sum_{\bf
 k'}\biggl\{\left[\left(\gamma_{\bf k-k'}+\gamma_{\bf
 k}\right)\sqrt[4]{\frac{1-\gamma_{\bf k'}}{\Delta +1+\gamma_{\bf
 k'}}}+\left(\gamma_{\bf k-k'}-\gamma_{\bf
 k}\right)\sqrt[4]{\frac{1+\gamma_{\bf k'}}{\Delta +1-\gamma_{\bf
 k'}}}\right]^2\nonumber\\
&\times&\left[1+n_B(\omega_{\bf k'})\right]A({\bf
 k-k'},\omega-\omega_{\bf k'})\nonumber\\
&+&\left[\left(\gamma_{\bf k-k'}+\gamma_{\bf
 k}\right)\sqrt[4]{\frac{1-\gamma_{\bf k'}}{\Delta +1+\gamma_{\bf
 k'}}}-\left(\gamma_{\bf k-k'}-\gamma_{\bf
 k}\right)\sqrt[4]{\frac{1+\gamma_{\bf k'}}{\Delta +1-\gamma_{\bf
 k'}}}\right]^2\nonumber\\
&\times&n_B(\omega_{\bf k'})A({\bf k-k'},\omega+\omega_{\bf k'})
 \biggr\}.
\label{ses}\end{eqnarray}
Excluding the numeric prefactor and some other small details this
formula is similar to the respective formula of the spin-wave
approximation. \cite{sherman98}

To check the validity of the approximations made we compare the
hole spectral function calculated using Eqs.~(\ref{hgf}) and
(\ref{ses}) for the case of one hole in a 4$\times$4 lattice with
the available exact-diagonalization data obtained in this system.
\cite{dagotto} The left panels in Fig.~\ref{figii} demonstrate the
results of the exact diagonalization, the right panels present our
calculations. Both series of calculations were performed for the
same set of parameters: $J/t=0.2$, $t<0$, $T=0$ and $\eta=0.1t$
($i\eta$ was added to the frequency $\omega$ in the denominator of
Green's function to visualize $\delta$-functions; in our
calculations parameters $C_1$ and $\alpha$ were estimated for low
but finite temperature $T=0.02t$). As can be seen from the figure,
the spectral functions obtained in our calculations are in good
agreement with the functions found in the exact diagonalization.
This agreement is somewhat better than that achieved in the
spin-wave approximation, \cite{marsiglio} because in contrast to
this approximation our approach takes into account the difference
between the spectral functions for wave vectors separated by
$(\pi,\pi)$ [cf.\ the spectra for ${\bf k}=(0,0)$ and
$(\pi,\pi)$]. As can be seen, in our approximation the binding
energy of the quasiparticle peak is underestimated in comparison
with the exact-diagonalization result. This may be connected with
the fact that the considered hole concentration $x=1/16$ is not
low enough and can lead to some deviations from the used spin
spectral function (\ref{ssf}).

An example of the one-hole zero-temperature spectra in a larger lattice
is given in Fig.~\ref{figiii}. We notice that the shapes of the spectra
cease to change perceptibly with increasing lattice size starting from
a 16$\times$16 lattice. Excluding the mentioned difference in spectra
with wave vectors spaced by $(\pi,\pi)$ they are close to those
obtained in the spin-wave approximation. \cite{marsiglio}

With the use of the Hubbard operators equations similar to
Eq.~(\ref{se}) were obtained also in Ref.~\onlinecite{plakida99} for
the two-dimensional $t$-$J$ model and in Ref.~\onlinecite{barabanov}
for a somewhat different model of the CuO$_2$ plane. The interaction
constant derived in the former work differs from the constant in
Eq.~(\ref{se}). The constant of Ref.~\onlinecite{plakida99} is not
applicable for low hole concentration: the spectral functions
calculated with it differ essentially from those obtained by the exact
diagonalization \cite{dagotto} and in the spin-wave approximation.
\cite{marsiglio,plakida94} However, this constant can be applicable in
the region of heavy doping.

\section{Concluding remarks}
Equation~(\ref{se}) was obtained under the supposition of a small hole
concentration. We have verified that in this limit Eq.~(\ref{se}) in
combination with Eq.~(\ref{hgf}) describes the hole spectral function
in good agreement with the exact diagonalization data. On the other
hand, with increasing hole concentration spin correlations are
weakened, the self-energy becomes small and elementary excitations
described by the two equations tend to the weakly correlated
nearest-neighbor band with the dispersion $\varepsilon_{\bf k}$,
Eq.~(\ref{urf}), where $C_1\rightarrow 0$. Thus, the obtained equations
give the correct behavior of the hole spectrum in the two limiting
cases. Besides, it was demonstrated that Eqs.~(\ref{sgf}) and
(\ref{freq}) with the parameters determined self-consistently give a
quantitatively correct description of the spin subsystem in the undoped
case. Equations of the spin-wave approximation, which are similar to
Eqs.~(\ref{sgf}), (\ref{freq}) and (\ref{po}), describe the rapid
weakening of spin correlations with hole doping, \cite{sherman98} as it
is necessary for the above-discussed transformation of the hole
spectrum from light to heavy doping. This gives ground to suppose that
the obtained equations can provide a qualitatively correct
interpolation between these two limiting cases.

\acknowledgements This work was partially supported by the ESF grant
No.~4022 and by the WTZ grant (Project EST-003-98) of the BMBF\@. A.S.
thanks International University Bremen for hospitality.

\begin{figure}\caption{The spin correlations ${\cal C}(l)=4|\langle
s^z_{\bf l}s^z_{\bf 0}\rangle|$, ${\bf l}=(l,0)$ calculated for
$T/J=0.5$, 0.75 and 1 in this work (open circles) and by the Monte
Carlo method in Ref.~\protect\onlinecite{makivic} (filled circles). In
both calculations a 32$\times$32 lattice without holes was used.}
\label{figi}\end{figure}

\begin{figure}\caption{The hole spectral function $A({\bf k}\omega)$
for the case of one hole in a 4$\times$4 lattice and parameters
$J=0.2t$, $\eta=0.1t$, and $T=0$. Left panels: exact-diagonalization
data from Ref.~\protect\onlinecite{dagotto}, right panels: our
calculations. The respective wave vectors are indicated in the upper
right corners of the panels.} \label{figii}\end{figure}

\begin{figure}\caption{The hole spectral function $A({\bf k}\omega)$
for the case of one hole in a 20$\times$20 lattice and parameters
$J=0.2t$, $\eta=0.01t$, and $T=0$. Wave vectors indicated near the
curves are selected along the symmetry lines $(0,0)-(0,\pi)$ in (a) and
$(0,0)-(\pi,\pi)$ in (b).} \label{figiii}\end{figure}

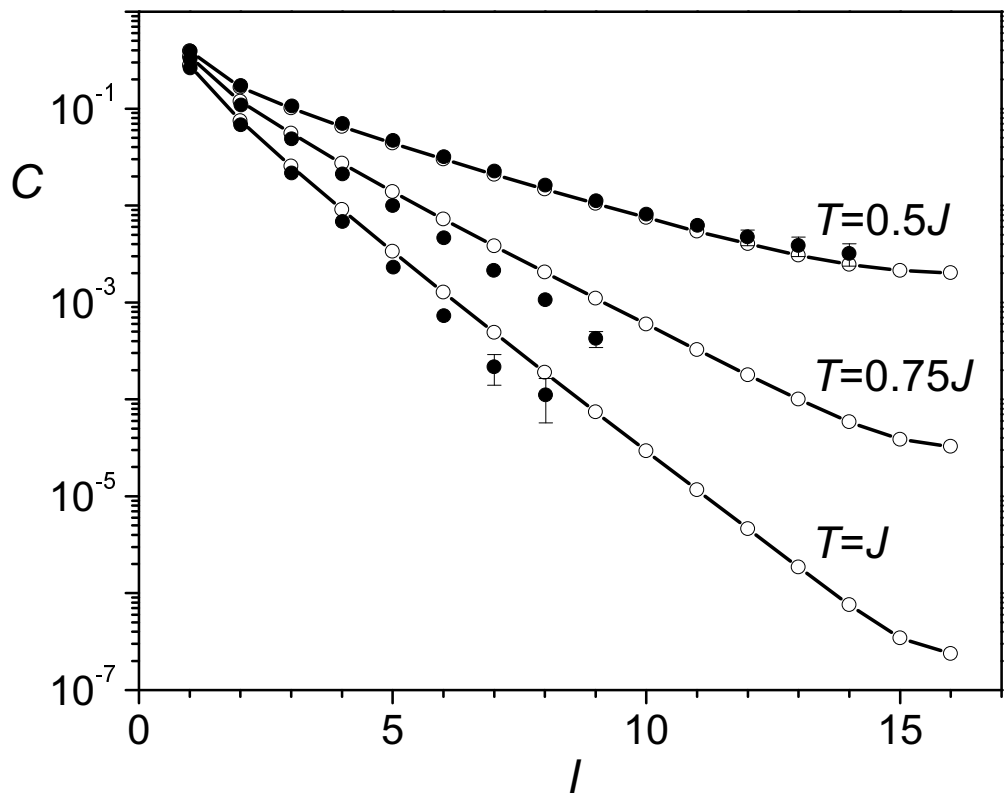
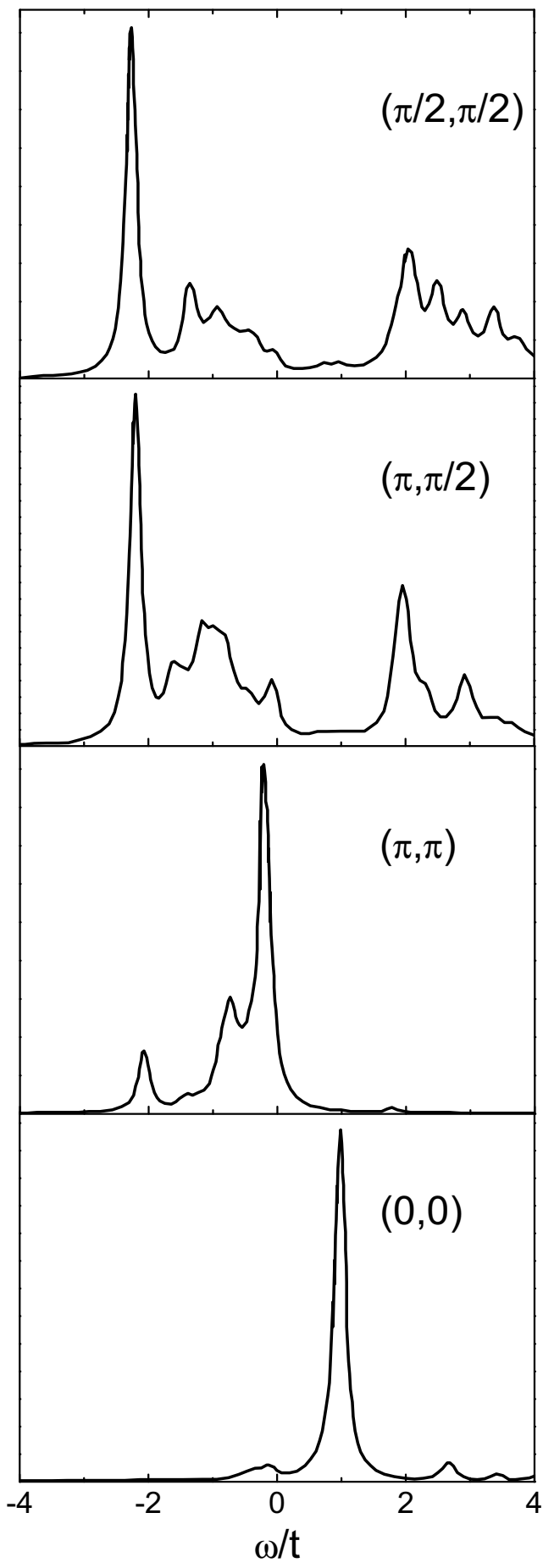
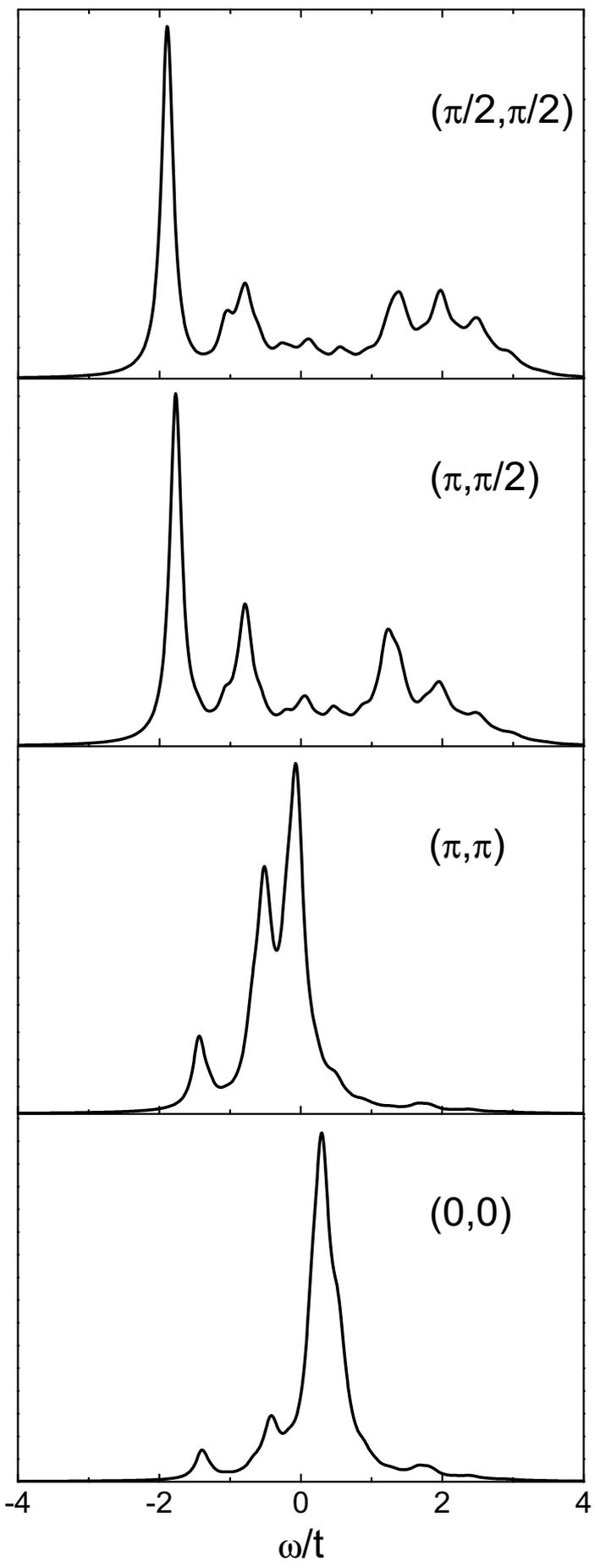
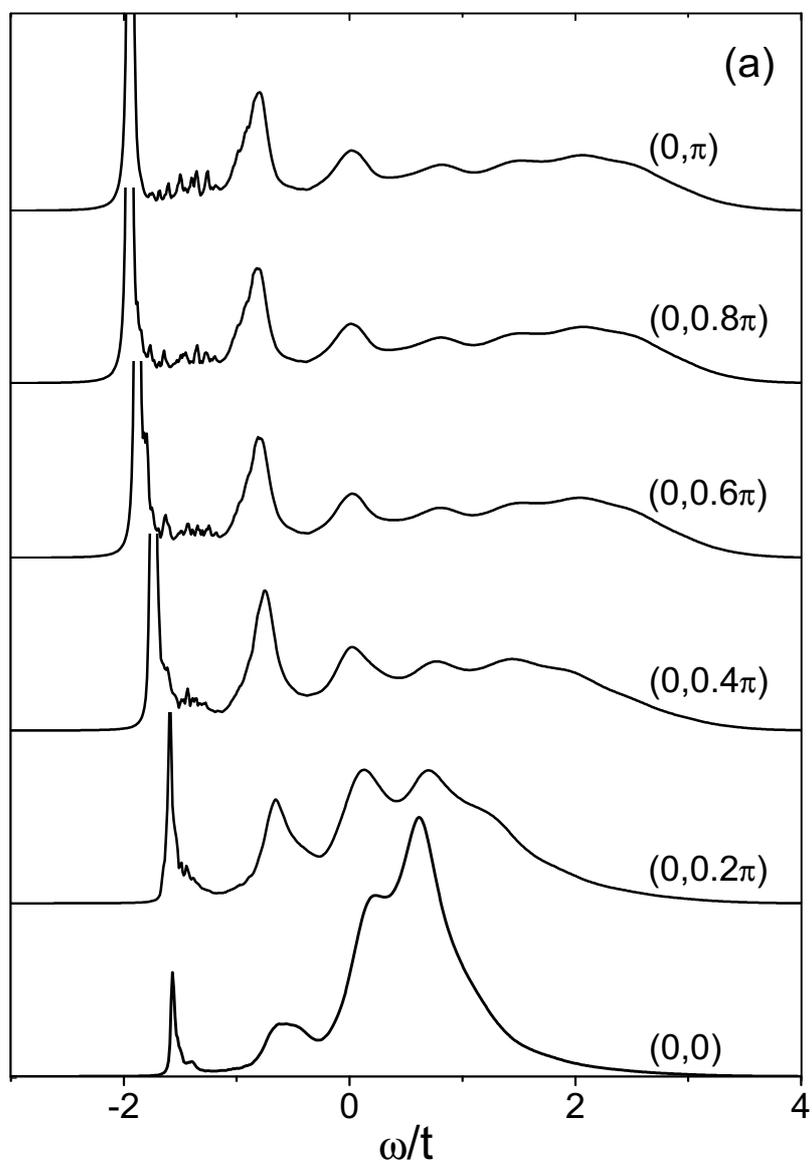
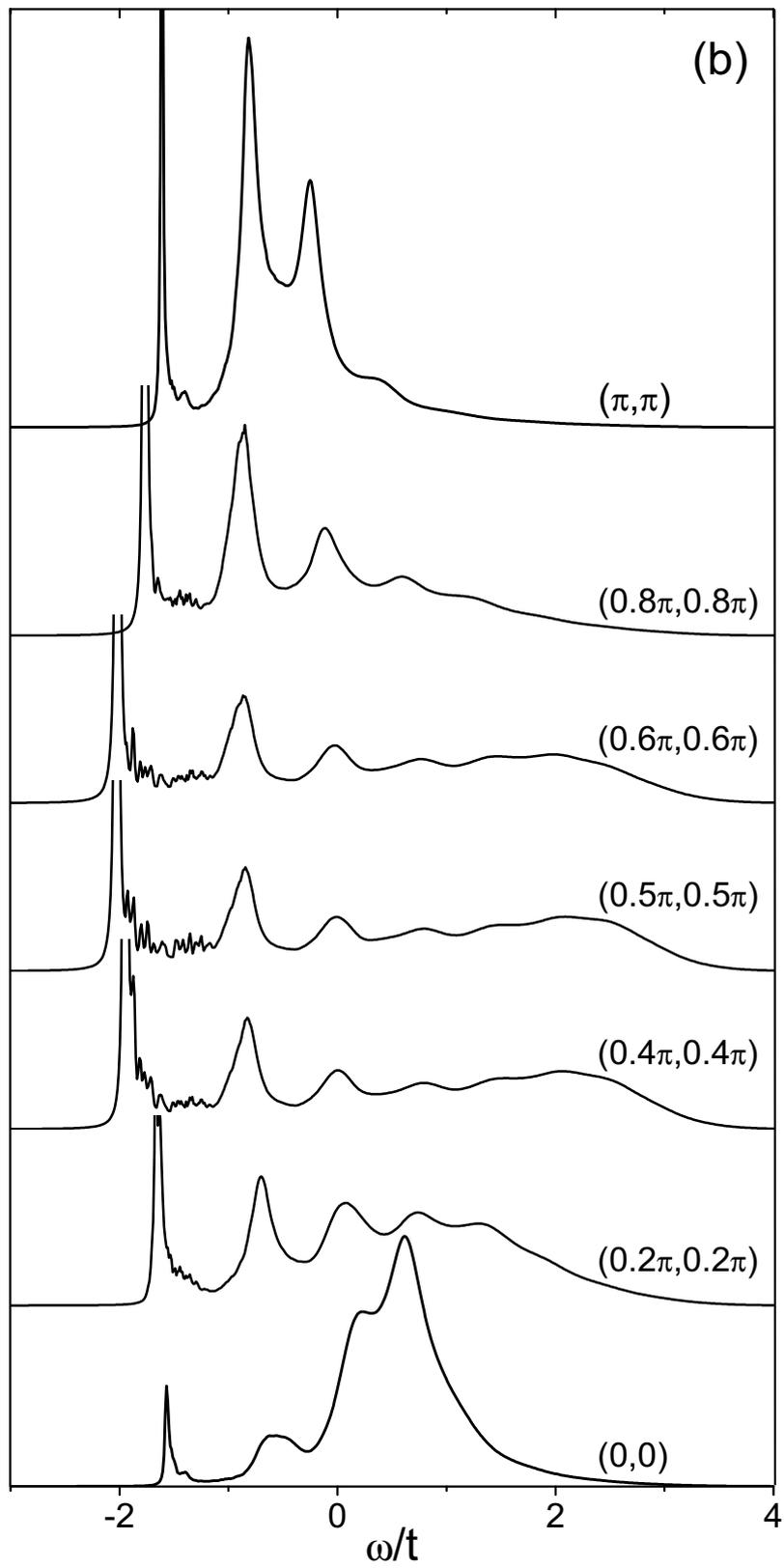
\end{document}